\documentclass[aps,prl,twocolumn,groupedaddress]{revtex4-1}

\usepackage{graphicx}
\usepackage{mathtools}
\usepackage{amssymb,amsmath}
\usepackage[caption=false]{subfig}
\usepackage{placeins}
\usepackage{verbatim}
\usepackage{natbib}

\begin{document}


\title{Forecasting Fluid Flows Using the Geometry of Turbulence}


\author{Balachandra Suri}
\thanks{These authors contributed equally to this work.}
\author{Jeffrey Tithof}
\thanks{These authors contributed equally to this work.}
\author{Roman O. Grigoriev}
\author{Michael  F. Schatz}
\email[]{michael.schatz@physics.gatech.edu}
\affiliation{School of Physics, Georgia Institute of Technology, Atlanta, Georgia 30332-0430, USA}


\date{\today}

\begin{abstract}

The existence and dynamical role of particular unstable Navier-Stokes 
solutions (exact coherent structures) is revealed in laboratory studies of weak turbulence in a thin, electromagnetically-driven fluid layer.   
We find that the dynamics exhibit clear signatures of numerous 
unstable equilibrium solutions, which are computed using a combination of flow measurements from the experiment and fully-resolved numerical simulations. We demonstrate the dynamical importance of these solutions by showing that turbulent flows visit their state space neighborhoods repeatedly.
Furthermore, we find that the unstable manifold associated with one such unstable equilibrium
predicts the evolution of turbulent flow in both experiment and simulation for a considerable period of time. 

\end{abstract}

\maketitle

Recent theoretical \cite{nagata_1990, kawahara_2001, Visw07b, gibson_2008, cvitanovic_2010, waleffe_2001, itano_2001, faisst_2003, duguet_2008a, willis_2013} and experimental \cite{hof_2004, lozar_2012, dennis_2014, lemoult_2014} studies have approached turbulence from a new perspective, which stems from the observation that turbulent flows often exhibit recognizable transient \textit{coherent} patterns that recur over time and space. 
These patterns are related to a class of unstable \textit{nonchaotic} solutions of the Navier-Stokes equation, called ``exact coherent structures" (ECS). 
Numerical studies have shown that ECS can serve as building blocks in describing both the statistical  \cite{kawahara_2001, chandler_2013} and dynamical \cite{gibson_2008, duguet_2008a, cvitanovic_2010} behavior of fluid turbulence at transitional Reynolds numbers. 

\begin{figure}
\centering
{\includegraphics[width=3in]{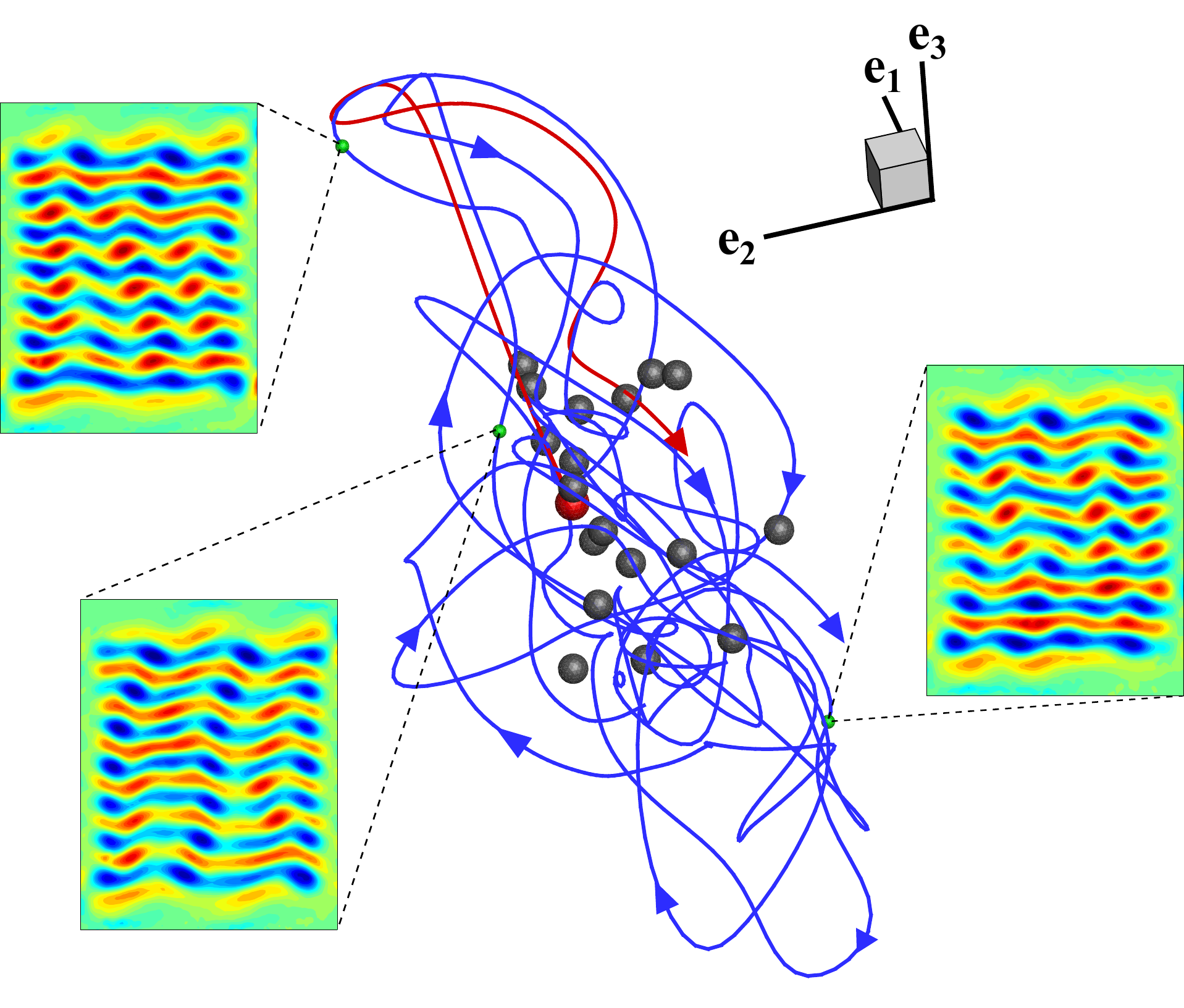}}
\caption{Chaotic evolution of a turbulent flow in the state space (blue trajectory in the middle) and in the physical space (the three snapshots of the vorticity of the flow reconstructed from experimental measurements). Each snapshot corresponds to a point (green dot) on the state space trajectory. Also shown are the equilibrium ECS (gray and red spheres) and the dominant unstable submanifold (red curve) associated with one of them. A three dimensional projection of the full ($O(10^5)$-dimensional) state space is shown.}
\label{fig:hairball}
\end{figure}

Deep physical insight emerges from a geometrical description of turbulence \cite{hopf_1948} in which the flow field at each instant is represented as a single point in a high-dimensional ($\gtrsim 10^5$) state space (Fig. \ref{fig:hairball}). 
The evolution of turbulent flow in the physical space, then, corresponds to movement in state space from one location to another along a tortuous trajectory. 
The ECS exist in the same regions of state space explored by turbulence, but being unstable, they are observed only fleetingly (Fig. \ref{fig:hairball}). 
When the turbulent trajectory passes through the neighborhood of an ECS, the stable and unstable manifolds of the ECS are expected to guide its approach and departure. 
Furthermore, the dynamical connections formed by intersections of stable and unstable manifolds of different ECS are expected to guide the turbulent trajectory from the vicinity of one ECS to another. Consequently, identifying the frequently visited ECS and the geometry of state space should enable a deterministic, low-dimensional description of fluid turbulence, at least near the transition from laminar to turbulent flow.

Few prior studies have tested this geometrical description experimentally. 
The vast majority of earlier research has focused on finding ECS in simulations of three-dimensional (3D) shear flows, such as plane Couette flow \cite{nagata_1990, kawahara_2001, Visw07b, gibson_2008, cvitanovic_2010}, plane Poiseuille flow \cite{waleffe_2001, itano_2001}, and pipe flow \cite{faisst_2003, duguet_2008a, willis_2013} in small boxes with idealized (periodic) boundary conditions. 
The few studies aimed at detecting ECS in experiments \cite{hof_2004, lozar_2012, dennis_2014, lemoult_2014} were all conducted using open flows, with the structures of interest advected past the imaging system and Taylor's hypothesis used to reconstruct their spatial structure. This, as well as the difference in the boundary conditions, only allows indirect comparison between experiment and theory.

In this article, we provide unambiguous experimental evidence for the dynamical relevance of ECS by studying turbulent flow generated in an electromagnetically forced shallow electrolyte layer. 
In particular, we focus on the dynamical role of unstable equilibria and provide the first experimental demonstration of a turbulent flow following the unstable manifold of a particular ECS.

\begin{figure}
\centering
\subfloat[]{\includegraphics[height=2.0in]{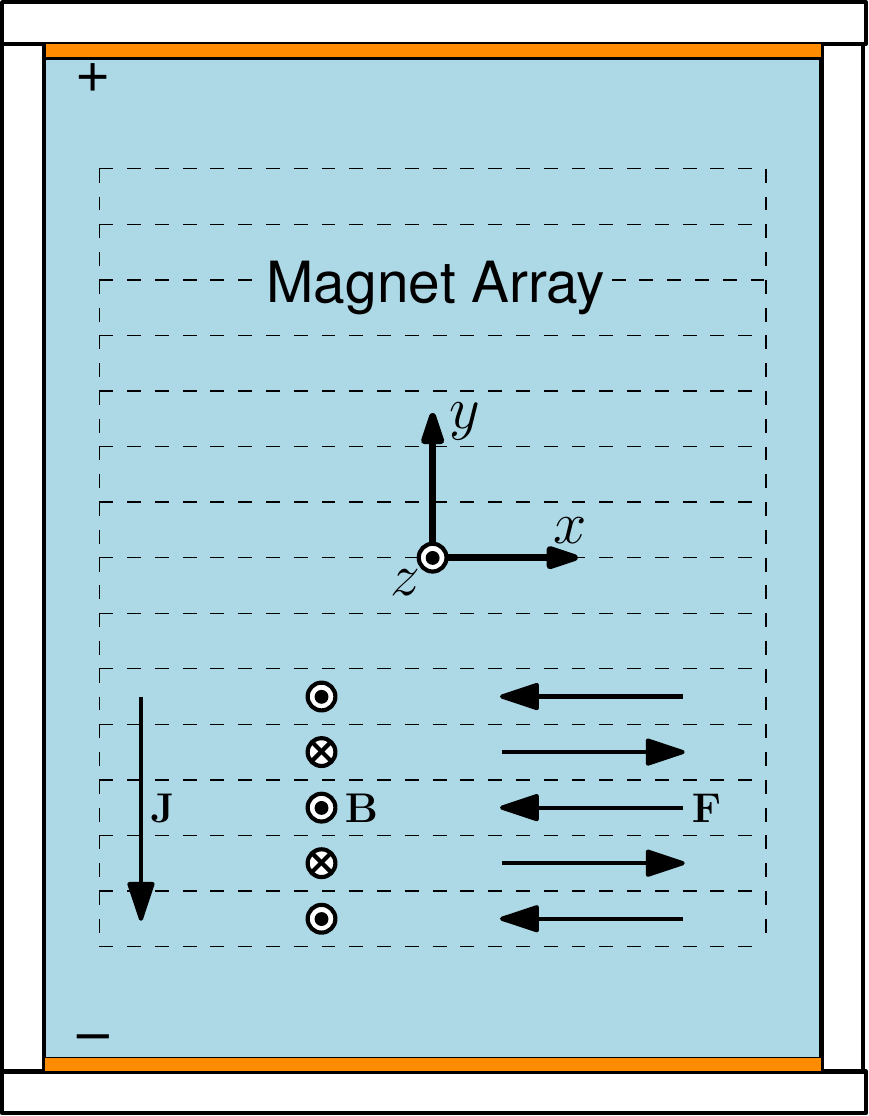}} \hspace{1mm}
\subfloat[]{\includegraphics[height=2.0in]{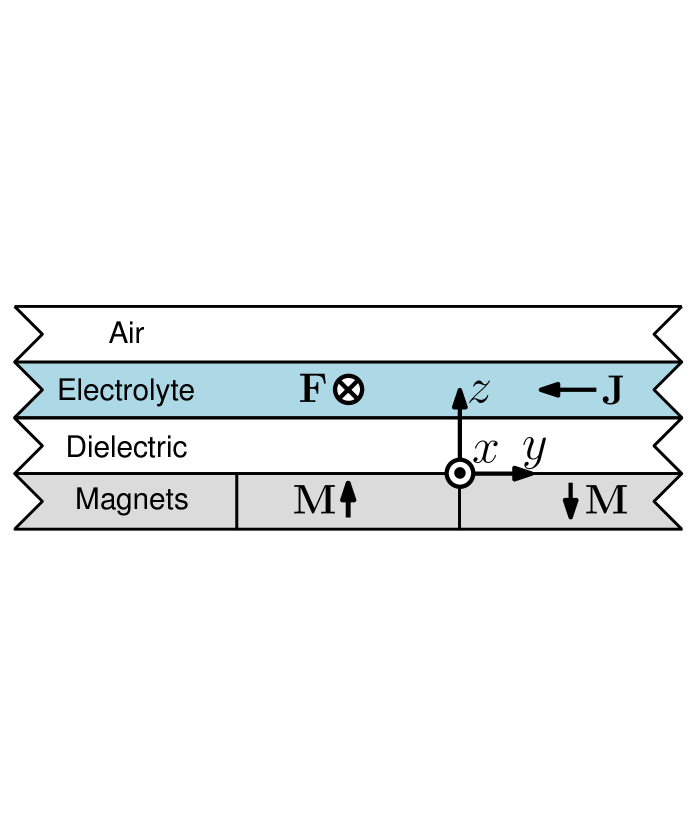}}
\caption{A schematic diagram of the experimental setup viewed from (a) above and (b) the side. In (a), dashed lines indicate the 14 permanent magnets forming the magnet array.
A direct current density $\bf J$ is passed through the electrolyte via two electrodes (orange rectangles), and the interaction with the magnetic field forces the fluid in the $\pm x$-direction.  
The direction of the current density $\bf J$, the magnetic field $\bf B$, and the fluid forcing ${\bf F} = {\bf J} \times {\bf B}$ are indicated over only the bottom 5 magnets in (a); in (b), {\bf M} indicates the magnetization.} \label{fig:exp_setup}
\end{figure}

Electromagnetically driven flows in a shallow electrolyte layer with lateral dimensions (in the $xy$-plane) much larger than the layer thickness (along $z$) have been previously studied extensively as models of atmospheric and oceanic flows \cite{dolzhansky_2012}. Such flows, often termed ``quasi-two-dimensional" (Q2D), can be treated as two-dimensional (2D) from both experimental and computational standpoints, which facilitates longer observation times.
The flow in our experiment is generated using two immiscible layers of fluid  \cite{suri_2014} in a 17.8 cm $\times$ 22.9 cm container. 
The top layer is an electrolyte and the bottom layer is a dielectric, each 3 mm thick. An array of long permanent magnets is placed below the fluid layers.  
Adjacent magnets in the array have alternating magnetization ${\bf M}$ in the $\pm z$-direction. A direct current with density ${\bf J} = J\hat{\bf y}$ passing through the electrolyte layer interacts with the magnetic field, producing a Lorentz force $\bf F = \bf J \times \bf B$. 
This force is in the $\pm x$-direction and approximately sinusoidal along the $y$-direction (with period 2.54 cm). 
We choose this form of forcing to approximate the canonical 2D Kolmogorov flow \cite{arnold_1960, chandler_2013}. 
The directions of $\bf J$, $\bf B$, and $\bf F$ are indicated in Fig. \ref{fig:exp_setup}.  
We image the flow and perform particle image velocimetry \cite{prana} to obtain temporally-resolved 2D velocity fields at the free surface of the fluid over the entire flow domain. 

The evolution of Q2D experimental flow is described using a 2D model \cite{suri_2014}, obtained by averaging the 3D Navier-Stokes equation over the confined direction ($z$):
\begin{equation}\label{eq:q2dns}
\frac{\partial {\bf u}}{\partial t} + \beta {\bf u}\cdot\nabla{\bf u} = -\nabla p +\bar{\nu}\nabla^2 {\bf u} - \alpha {\bf u}+ \langle {\bf F} \rangle_z. \\
\end{equation}
Here, the incompressible 2D velocity field ${\bf u}(x,y,t)$ represents the flow at the electrolyte-air interface in the experiment. Also, pressure $p$ and the depth-averaged forcing $\langle {\bf F}\rangle_z$ depend only on the coordinates in the extended directions $(x,y)$. The constants $\beta$ and $\alpha$ arise as a result of depth-averaging and capture the effects due to the solid boundary at the bottom, and  $\bar{\nu}$ is the depth-averaged viscosity \citep{suri_2014}.
For the experimental setup described above, we obtain $\beta = 0.83$, $\alpha = 0.64$ s$^{-1}$,  and $\bar{\nu} = 3.23 \times 10^{-6}$ m$^2$/s. 
Equation (\ref{eq:q2dns}) is numerically solved in the primitive variable (${\bf u}, p$) formulation using a fractional-step finite difference projection method \cite{griebel_1998,armfield_1999} on a domain with lateral dimensions identical to the experiment. To facilitate direct quantitative comparison with experiment we impose realistic no-slip boundary conditions in the lateral directions and employ an experimentally-validated forcing profile \cite{tithof_2016}.

In both experiment and simulation, we characterize the complexity of the dynamics using the Reynolds number $Re = U L/{\bar{\nu}}$; here, $U$ is the temporal average of the spatial root mean square velocity and $L=1.27$ cm is the magnet width.
In the present experimental setup, as the current density is increased, the flow transitions through a series of bifurcations, and eventually becomes weakly turbulent at $Re\approx18$ \cite{tithof_2016}; our analysis herein is performed at $Re=22.5$. 
Turbulence arises in 2D flows at Reynolds numbers $Re$ that are much lower than in 3D flows. In particular, turbulent cascades with 
characteristic scaling behaviors arise in 2D for $Re$  as small as a few hundred \cite{boffetta_2012}; in 3D, many flows are still laminar in the same range of Reynolds numbers. 

It is unknown {\it a priori} whether the evolution equation (\ref{eq:q2dns}), for the values of $Re$ describing the experiment, possesses any unstable nonchaotic solutions  or whether any of these solutions play a dynamically important role in the evolution of the turbulent flow.
However, the turbulent flow fields can be tested for signatures of various types of ECS. For instance, we found that the evolution of the flow in the experiment is punctuated by brief intervals when the velocity field ${\bf u}$ becomes almost stationary. Correspondingly, there are deep minima in the rate at which the flow changes, which we quantify using:
\begin{equation}\label{eq:s}
s(t)=\left\|\frac{\partial{\bf u}}{\partial t} \right\| 
\approx \frac{1}{\Delta t}\left\{\iint\left[{\bf u}(t+\Delta t) - {\bf u}(t)\right]^2 \,dx\,dy\right\}^{1/2},
\end{equation}
where the integral is taken over the entire spatial domain and $\Delta t$ defines the temporal frequency at which velocity fields are sampled in experiments. These slow-downs in the evolution suggest that the turbulent trajectory passes near an unstable equilibrium 
{(i.e., a ``fixed point")}, 
where $s(t)=0$, just like an inverted pendulum that slows down when it passes through the unstable equilibrium at the top of its swing.

\begin{figure}
\centering
\subfloat[]{\includegraphics[height=1.04in]{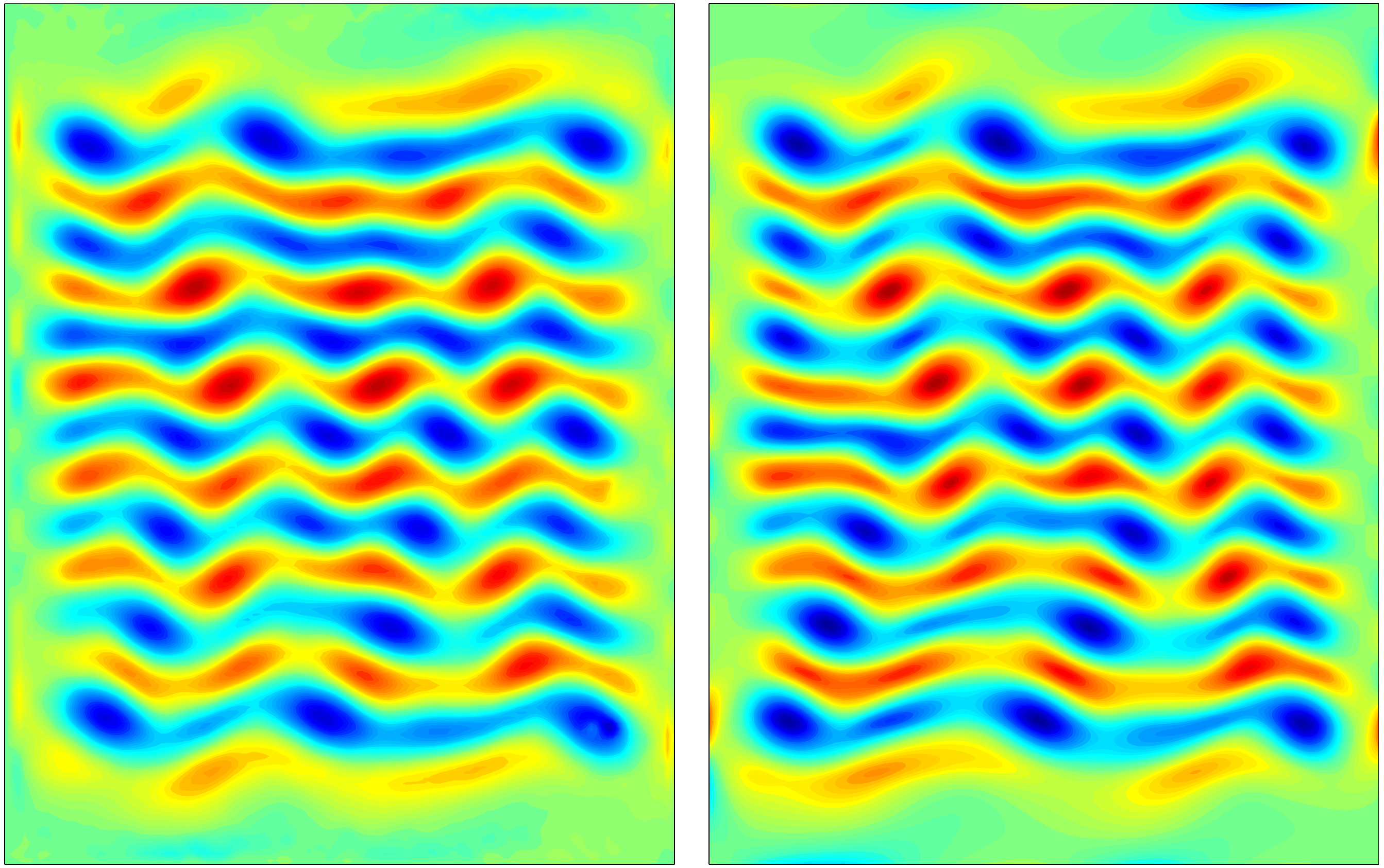}} \hspace{1mm}
\subfloat[]{\includegraphics[height=1.04in]{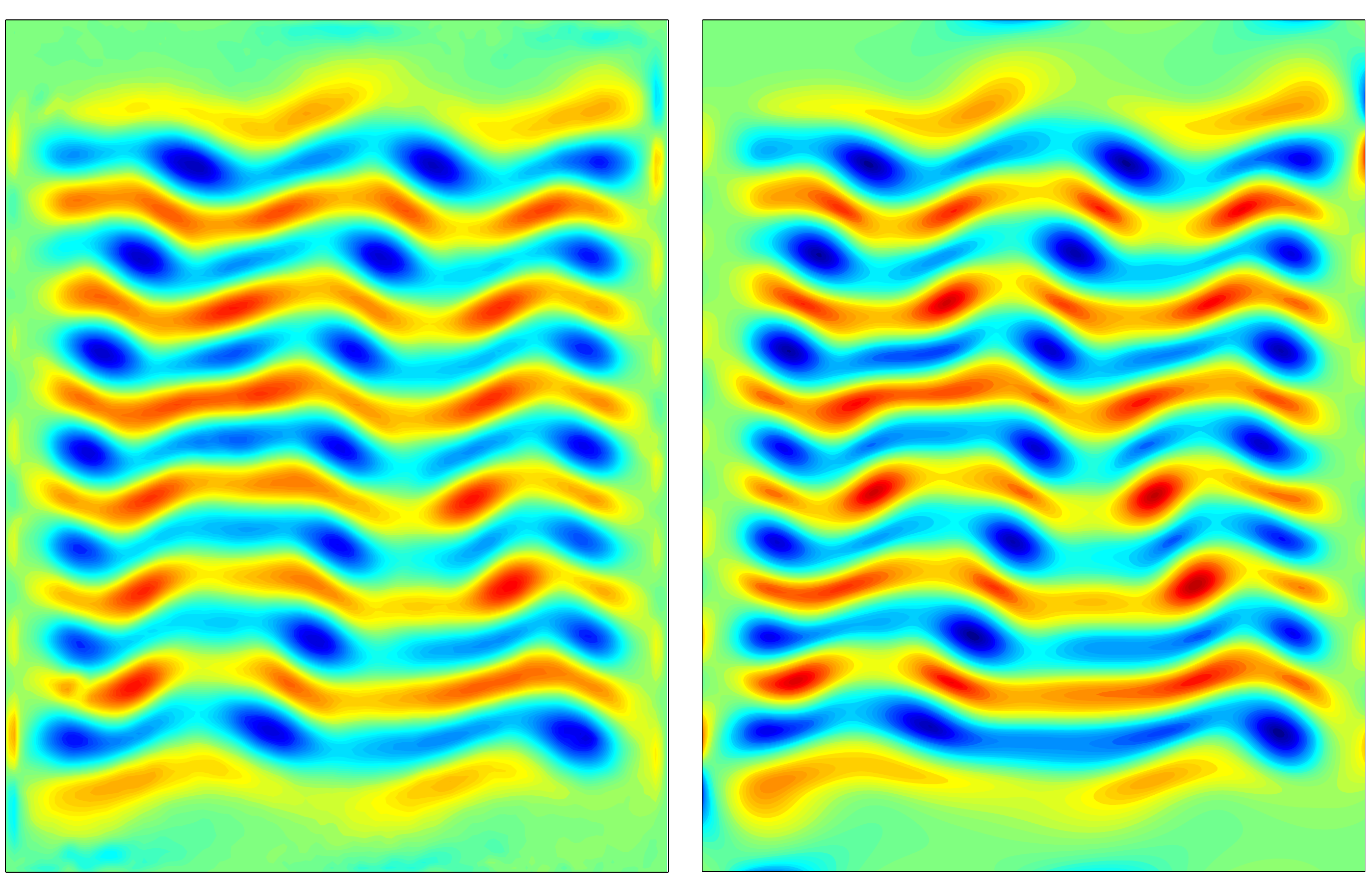}}
\caption{Two experimental states at instants of dramatic slow-down in the evolution (left image in (a) and (b)) 
 and the corresponding unstable equilibria computed using a Newton solver (right image in (a) and (b)). The colormap shows the vorticity, $\omega=(\nabla\times{\bf u})\cdot{\hat z}$ (see Fig. 4(b)).}
\label{fig:flow_fields}
\end{figure}

Moreover, the spatial structure of the unstable equilibrium should be similar to the turbulent flow field at the instant when the  evolution slows down. Using experimental flow fields ${\bf u}(t)$ from such instants as initial conditions, we can compute the corresponding unstable equilibria of equation (\ref{eq:q2dns}) using a matrix-free iterative Newton solver \cite{kelley_2003}.
As an illustration, Fig. \ref{fig:flow_fields} shows two examples of experimental flow fields used as initial conditions (left image in (a) and (b)) and the corresponding unstable equilibria computed (right image in (a) and (b)).  The degree of similarity is striking, suggesting that turbulent trajectories pass quite close to the ECS.

In all, we have computed 13 distinct solutions initializing the Newton solver with flow fields from the experiment.  Using an identical methodology, we have also computed 10 distinct unstable equilibrium solutions using initial guesses from a numerically generated turbulent trajectory. 
Of these, 4 were found to coincide with the ones computed from experimental initial conditions. All of the computed ECS are shown as (gray or red) spheres in Fig. \ref{fig:hairball}. 
The frequency with which the neighborhoods of the unstable equilibria are visited by the turbulent flow (blue curve) suggests that these ECS play an important dynamical role.

The dynamics in the neighborhood of an ECS are conjectured to be guided by the stable and unstable manifolds of that ECS: Turbulent trajectories should approach an ECS following its stable manifold and recede following its unstable manifold. 
However, computing high-dimensional stable manifolds is substantially more difficult than computing relatively low-dimensional unstable manifolds; hence, we focus our study on demonstrating the dynamical role of ECS by showing that turbulent trajectories departing from the neighborhood of an ECS indeed follow its unstable manifold. 
Consider, for example, the unstable equilibrium  shown in Fig. \ref{fig:flow_fields}(b). 
This equilibrium has seven unstable eigendirections, with the associated eigenvalues given by 0.1492, 0.0147 $\pm$ 0.1680$i$, 0.0045 $\pm$ 0.1104$i$, and 0.0009 $\pm$ 0.4500$i$. 
As can be seen, these eigenvalues show a clear hierarchy, with the leading (real) one being ten times larger than the real part of the remaining six. 
Hence, as they recede from the ECS, we expect that turbulent trajectories should be guided primarily by the one-dimensional (1D) invariant submanifold, which corresponds to a trajectory that starts at the equilibrium and evolves in the direction of the leading unstable eigenvector.

\begin{figure}
\centering
\subfloat[]{\includegraphics[height=1.75in, angle = 0]{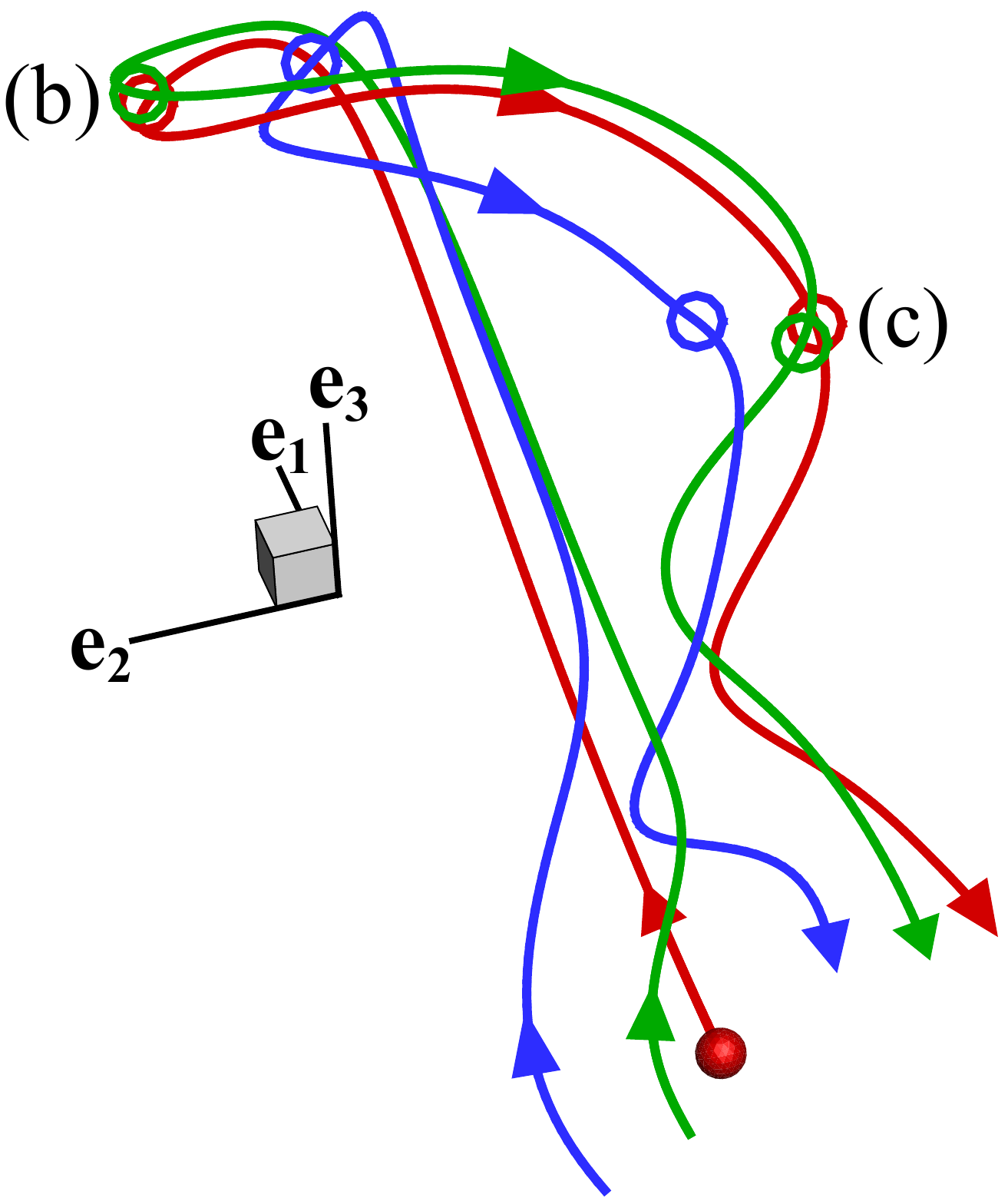}}\\
\subfloat[]{\includegraphics[height=1.19in]{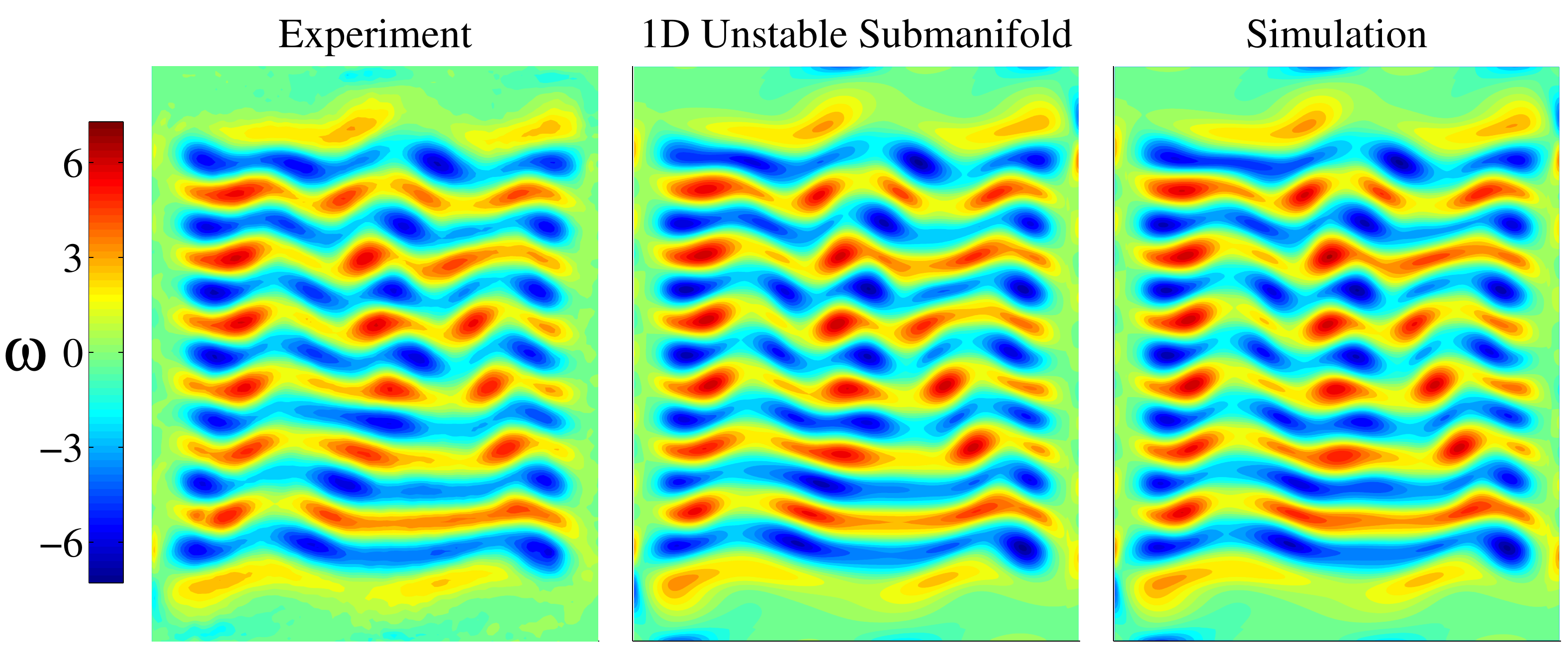}\hspace{0.24in}}\\
\subfloat[]{\includegraphics[height=1.16in]{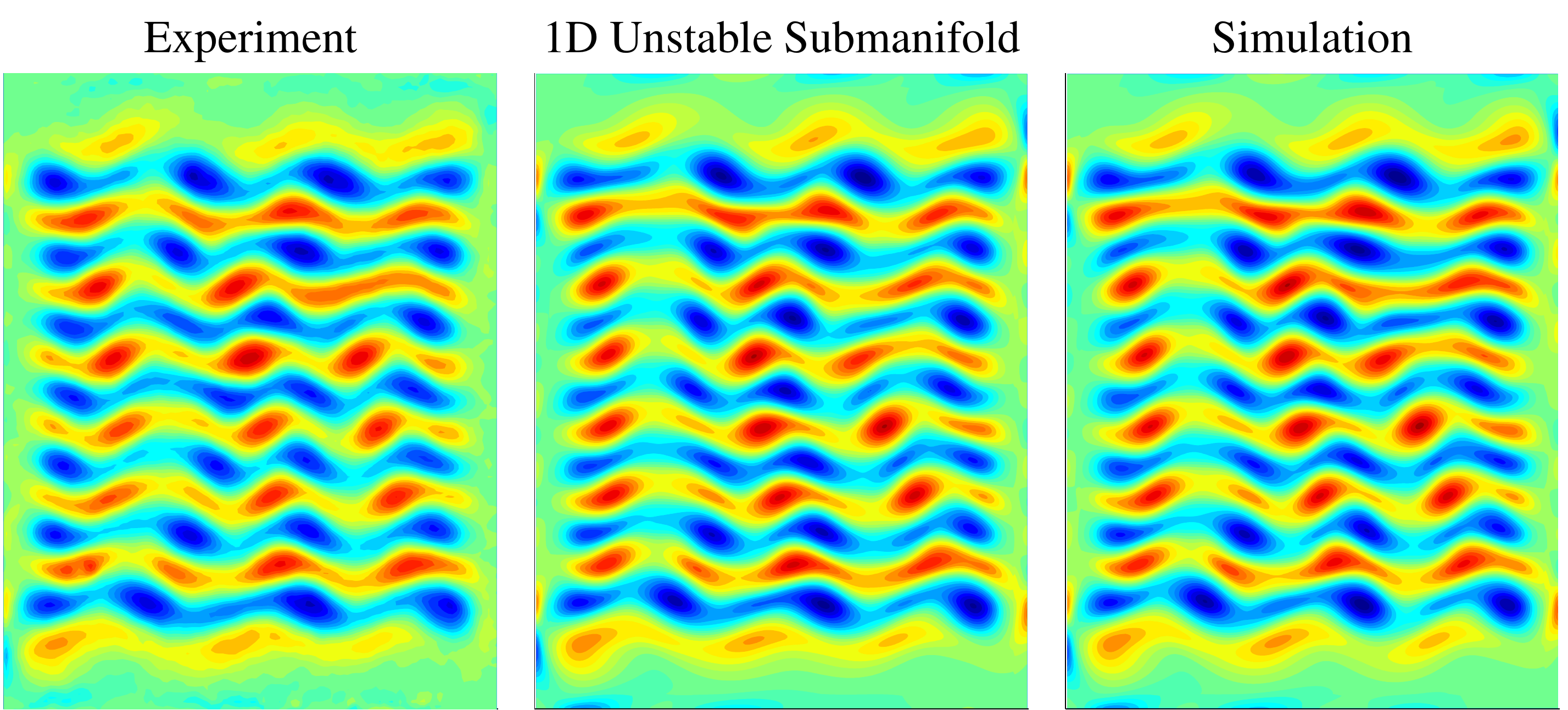}}
\caption{(a) A projection showing an unstable equilibrium (red sphere) with the dominant unstable submanifold (red curve). Both the experimental (blue curve) and numerical (green curve) turbulent trajectories follow this submanifold as they depart from the neighborhood of the unstable equilibrium. (b-c) Representative flow fields from the submanifold and two turbulent trajectories at two different time instances (denoted with circles in (a)).}
\label{fig:projections}
\end{figure}

To visualize this, we project the state space trajectories onto a subspace spanned by the dominant eigenvector ${\bf e}_1$ and vectors ${\bf e}_2$ and ${\bf e}_3$ constructed from the eigenvectors associated with a complex conjugate pair such that ${\bf e}_1$, ${\bf e}_2$, and ${\bf e}_3$ form an orthogonal basis.
The projection is shown in Figs. \ref{fig:hairball} and \ref{fig:projections}(a) where the red sphere and the red curve denote the unstable equilibrium and its dominant unstable submanifold, respectively. 
Note that the trajectory which defines the submanifold is straight only close to the ECS and becomes curved further away due to the nonlinearity of the evolution equation (\ref{eq:q2dns}). 
The blue and green curves in Fig. \ref{fig:projections}(a) correspond to the experimental and numerical turbulent trajectories receding from the ECS. 
They follow the submanifold remarkably well even far away from the ECS (cf. Fig. \ref{fig:hairball}), with the length of the experimental curve corresponding to a total of 3.4 correlation times.
In the immediate vicinity of the ECS the dynamics are well-approximated by the linearization of the evolution equation \eqref{eq:q2dns}, which ensures that the unstable manifold attracts neighboring trajectories. It should be emphasized, however, that the dominant unstable submanifold is {\it not} locally attractive, yet it guides neighboring trajectories quite far from the ECS, where linearization certainly becomes invalid.

To illustrate the degree of similarity between the turbulent trajectories and the dominant submanifold in the physical space, we compare snapshots of the flow fields in Figs. \ref{fig:projections}(b) and \ref{fig:projections}(c). The corresponding time instants are indicated by circles in Fig. \ref{fig:projections}(a). The flow fields corresponding to all three trajectories remain very similar at every time instance, validating the conjecture that dominant submanifolds guide neighboring turbulent trajectories.

Our findings demonstrate the power of this geometrical approach: Using direct numerical simulations one can \textit{pre-compute} ECS and their dominant submanifolds, once and for all. 
Turbulent evolution can then be forecast moderately far into the future,  based solely on the fact that the turbulent trajectory was found close to an ECS. 
By computing a sufficiently large set of ECS, including time-periodic ones, and their dominant submanifolds, it should be possible to ``tile'' the entire region of state space inhabited by turbulence.  
Such tiling should also enable {\it long-term} predictions which requires that the chaotic trajectory be periodically refined by comparing predictions with measurements, as done currently in weather prediction.
How well this deterministic picture succeeds in describing turbulence at higher Reynolds numbers remains an open question, but that question is far outside the scope of this study.

This article provides the first direct and unambiguous evidence in support of the dynamical role played by exact coherent structures in turbulent fluid flows at transitional Reynolds numbers, in both numerics and experiment. 
In contrast to previous studies of ECS, we used a fully resolved model of the entire experimental system, including proper boundary conditions.
We used experimental flow fields to identify a large number of unstable equilibrium solutions of the depth-averaged Navier-Stokes equation and demonstrated that when the evolution slows down, as it should near an equilibrium, the experimental flow fields become almost indistinguishable from the corresponding unstable equilibria. 
We have also validated another theoretical prediction, that the evolution of turbulent flow in state space is guided by unstable manifolds of ECS. More specifically, the flow follows the dominant submanifold that corresponds to the most unstable mode of the associated ECS.

\begin{acknowledgments}
This work is supported by the National Science Foundation under grants CMMI-1234436 and DMS-1125302.
\end{acknowledgments}


%

\end{document}